\documentclass[aps,prl,twocolumn,showpacs,superscriptaddress,groupedaddress]{revtex4}  
\usepackage{graphicx}  
\usepackage{dcolumn}   
\usepackage{bm}        
\usepackage{amssymb}   

\begin{document}



\title{Quantum Limit on Stability of Clocks in a Gravitational Field}
\author{Supurna Sinha $^{}$\& Joseph Samuel $^{}$}
\address{Raman Research Institute, Bangalore 560080, India.}
\date{\today}

\begin{abstract}
Good clocks are of importance both to fundamental physics
and for applications in astronomy, metrology and global positioning 
systems.
In a recent technological breakthrough, researchers at NIST have been able 
to achieve a stability 
of 1 part in $10^{18}$ using an Ytterbium clock. This naturally raises the 
question of whether
there are fundamental limits to the stability of clocks.
In this paper we point out that gravity and quantum mechanics
set a fundamental limit on the stability of clocks.
This limit comes from a combination of the uncertainty relation, 
the gravitational redshift
and the relativistic time dilation effect.
For example, a single ion hydrogen maser clock in a terrestrial 
gravitational
field cannot achieve a stability better than one part in $10^{22}$.
This observation has implications for laboratory experiments
involving both gravity and quantum theory. 
\end{abstract}

\maketitle


Time is a key concept in physics. We use clocks to measure time.
The history of timekeeping \cite{davasobel,hist} has seen the development 
of two kinds of clocks--
terrestrial and celestial. The earliest clocks were celestial: the earth 
going around the sun, the moon
around the earth, and the earth rotating on its axis, which measure a 
year, a month and a day respectively.
Early examples of terrestrial clocks are the clepsydra or water clock, the 
hour glass and  the pendulum.
Clocks are dynamical systems which exihibit a periodic
(e.g pendula) or decaying (e.g radioactive decay) behavior. Recent
examples of terrestrial clocks are quartz oscillators and atomic clocks.
In the twentieth century, it was realized that the rotation of the earth 
is not quite uniform and that
astronomers and physicists need better clocks\cite{astro}
than the spinning earth.
Over the years terrestrial clocks have so improved that they now form the 
standard for timekeeping\cite{handbook}.
As first suggested by Kelvin \cite{kelvin} (Sir William Thomson) and
later in the context of magnetic resonance by I.I.
Rabi\cite{rabi}, atomic spectral lines provide a reliable standard for 
time and frequency.
Due to giant strides in cold atom physics, we now have atomic 
clocks\cite{wineland} 
which are accurate to one part in $10^{18}$.
Atomic clocks have led to innovations like global positioning systems, 
advanced communications, tests of general relativity and
of variation of fundamental constants in nature.

Space and time provide the arena in which physics occurs. Traditionally,
space intervals are determined by measuring rods and time intervals by clocks.
Both rods and clocks are material
objects, subject to physical laws. For example, it was earlier believed that
measuring rods could be taken to be rigid, a belief which had to be revised
with the advent of special relativity. With the realisation that the speed
of light is finite and a constant of nature, we could reduce length measurements
to time measurements and eliminate measuring rods in favor of clocks. Since
clocks also are material objects, subject to the laws of physics,
we must ask whether there are limits on their time keeping from fundamental
physics.  
We would like clocks to be both accurate and stable. Accurate clocks  will 
deliver a periodic signal close to
a fiducial frequency. Stable clocks reliably maintain the same periodic 
signal and do not wander in frequency.
Atomic clocks have now progressed to the point where 
they are both accurate and stable.
The current record\cite{nist} for stability is one part in $10^{18}$. The 
current obstacles to improving clocks
are technological in nature. 
Stray electric fields, uncertainty in the height of the trap, 
ambient black body radiation and
excess micromotion of the clock all have to be brought under control 
before further progress can be made.

Is there
a fundamental limit, set by the laws of physics, to the
stability or accuracy of clocks. 
This question was earlier raised by Salecker and Wigner\cite{wignerq,wignerrmp,barrow},
who derived theoretical limits on freely moving clocks.
They consider the recoil momentum of clocks and derive lower bounds on the
mass of a clock given a certain desired accuracy. Our analysis differs from 
theirs in two respects. First, we consider clocks in a trap 
as realized in present day laboratories 
and arrange the 
stiffness so that the recoil is absorbed by the trap as a whole (Lamb-Dicke 
regime); second we consider the effect on an external gravitational field
such as that of the earth. 
We address the issue of stability of 
clocks in a gravitational field.
We suppose that the technological
problems have been solved: stray electric fields have been eliminated, the 
clock has been cooled to absolute
zero and so on. With these technological problems out of the way, we ask 
if there is a fundamental limit
to the accuracy of clocks in a gravitational field. The answer it turns 
out is yes. The gravitational redshift
and the uncertainty principle do impose such a limit.

We will see below that gravity and quantum mechanics combine to limit the 
stability attainable by a clock
of a given mass $m$. The limit stems from an unavoidable uncertainty in 
the position and motion of
the center of mass of the clock. Atomic clocks are cooled to still their 
random thermal motion,
which would otherwise cause variations in the apparent clock rate due to 
the Doppler effect.
Even if a clock is cooled to absolute zero, quantum zero point motion 
still remains.
In an external gravitational field, there are uncertainties in the clock 
rate arising from both
positional and motional uncertainty of the clock and this gives rise to 
fundamental limitations.
Clock stabilities currently achieved in laboratories are several orders 
below the fundamental quantum
limit. However, the existence of a quantum limit to timekeeping
{\it is} important because of its fundamental significance. Our limit reveals 
low energy connections
between gravitation and quantum mechanics which could help understand the 
relation between the two theories.

Consider a clock of mass $m$ placed in an uniform gravitational field $g$.
For definiteness we will suppose the clock to be a single atom (possibly 
an ion)  in a trap,
but our considerations are general.
A free atom at rest emitting a photon of frequency $\nu$ suffers a recoil which
causes a first order  Doppler shift $\frac{\delta \nu}{\nu}\approx v/c$,
where $v$ is the recoil velocity. 
If $\nu$ is the frequency of the emitted photon, the recoil momentum is $\frac{h \nu}{c}$ and the recoil energy
is $E_r=\frac{1}{2m}(\frac{h \nu}{c})^2$. This recoil effect 
can be eliminated by using a stiff trap, whose 
energy spacings are much larger than $E_r$.

This is the Lamb-Dicke regime\cite{dicke,cohentannoud},
where the recoil of the photon is taken by the trap as a whole and not 
just the atom.

We suppose the $Z$ axis of our coordinate system is chosen ``vertical'' in 
the direction 
determined by the gravitational
field as ``up''.
Let the quantum uncertainty in the vertical
position of the atom be $\Delta Z$
and the uncertainty in the vertical momentum be $\Delta P_Z$.
In the equations below we drop
numerical factors of order one to avoid clutter.
As is well known, the rate of clocks is affected by a gravitational field.
If the location of the clock has an uncertainty $\Delta z$ in the vertical 
direction, the clock rate has an uncertainty given by 
\begin{equation}
\left(\frac{\Delta \nu}{\nu}\right)_g=\left(\frac{\Delta 
T}{T}\right)_g\approx\frac{g \Delta Z}{c^2}
\label{uncertaintyt}
\end{equation}
This is a famous observation that emerged during the Bohr-Einstein
debates at the Solvay conference\cite{solvay}.
See Ref.\cite{solvay} for a thought experiment which is used to derive 
Eq.(\ref{uncertaintyt}). 
It is convenient to introduce a natural length $L_g=\frac{c^2}{g}$.
For a terrestrial gravitational field of $10 \rm m/s^2$, $L_g$ works out 
to be $10^{16}\rm m$, which is
about a lightyear.
Thus we have
\begin{equation}
\left(\frac{\Delta T}{T}\right)_g\approx \frac{\Delta Z}{{L_g}}
\label{elguncertainty}
\end{equation}
One might expect that by making $\Delta Z$ small, the stability of time 
measurement can be made arbitrarily high.
However, because of Heisenberg's uncertainty relation
$\Delta Z \Delta P_Z \ge \hbar/2$,
confining the clock in a stiffer trap to reduce the position uncertainty 
results in a large momentum
uncertainty in the vertical $Z$ direction.
The momentum uncertainty in the vertical direction can be viewed as a 
random  motion of the clock.
We can  suppose that $ P_Z \approx \Delta P_Z$. Using the relations 
$P_Z=\gamma m v$
, where $\gamma$ is the Lorentz factor, which exceeds one by an amount of 
order $(v/c)^2$.
We can estimate the associated velocity of the random zero point vertical
motion of the clock. The special relativistic time dilation effect due to 
the random motion
causes uncertainty, which is of order
$v^2/c^2$.
The time dilation leads to an uncertainty
in the clock rate:
\begin{equation}
\left(\frac{\Delta T}{T}\right)_t\approx 
\frac{v^2}{c^2}=\frac{P_Z^2}{m^2c^2}
=\frac{\hbar^2}{m^2c^2(\Delta Z)^2}
\label{rel}
\end{equation}
Squeezing the atom in the vertical direction to minimise the gravitational 
uncertainty in the clock
rate results in increased uncertainty due to the time dilation effect. 
There is thus an optimal
value for the width of the trap and a corresponding limit to the stability 
of the clock.
These ideas can be made slightly more precise
for the case of a harmonic trap.
To make contact with current experimental efforts 
\cite{wineland,nuclearclocks}
to make better clocks, we
specialize these general considerations to the case of a harmonic trap. An
atom clock of mass $m$, in a gravitational field $g$ is trapped in a 
harmonic
trap with angular frequency $\Omega_t$. If the operating frequency of the
clock is $\nu= \omega_c/(2\pi)$, where $\omega_c$ is the angular 
frequency,
the Doppler recoil energy of the clock is 
$$E_r=\frac{1}{2m}\left(\frac{\hbar
\omega_c}{c}\right)^2.$$
To prevent this recoil from affecting the observed frequency of the clock
we choose $\Omega_t$ rather larger than $\Omega_{LD}=\frac{E_r}{\hbar}$
so that we are in the
Lamb-Dicke regime. The recoil momentum is then taken not by the atom but
by the whole trap. This is very similar to the M{\"o}ssbauer effect
\cite{mossbauer}.

For an atom whose centre of mass motion is
in the quantum ground state of a harmonic trap,
we have an expression for the variance in the height got by equating
the mean kinetic (or potential) energy to half the zero point energy.
$$\frac{1}{2}m \Omega_t^2{\Delta Z}^2= \frac{\hbar \Omega_t}{4}$$
 which gives
$${\Delta Z}^2= \frac{\hbar}{2m\Omega_t}.$$ This quantum uncertainty in 
the
vertical position of the clock gives an uncertainty in the clock rate.
Let us introduce $\Omega_g = \frac{g}{c}$, which has the dimensions of
frequency and write

\begin{equation}
(\frac{\delta \nu}{\nu})_g \approx \frac{g \Delta
Z}{c^2}=\frac{g}{c}\sqrt{\frac{\hbar}{2mc^2\Omega_t}}\approx\frac{\Omega_g}{\sqrt{ 
\omega_m
\Omega_t}}
\label{gravity}
\end{equation}
where $\omega_m=\frac{mc^2}{\hbar}$ is the mass of the clock converted 
into
frequency units. (Note that $\omega_m$ is $\underline{not}$ the operating
angular frequency of the clock, which we denote by $\omega_c=2\pi\nu$).

It is evident that the gravitational contribution is made smaller by
stiffening the trap. However, this leads to larger accelerations within
the trap. A typical acceleration is $a=\Omega_t^2 \Delta Z$ which gives 
rise to
a time dilation redshift of
\begin{equation}
(\frac{\delta \nu}{\nu})_{t} \approx \frac{a \Delta
Z}{c^2}\approx\frac{\Omega_t^2}{c^2}\frac{\hbar}{m\Omega_t}\approx\frac{\Omega_t}{\omega_m}.
\label{timedilation}
\end{equation}
As we know from the principle of equivalence, acceleration has the same 
effect
as gravity; hence the similarity between the first equalities of
(\ref{gravity}) and (\ref{timedilation}).

Adding the variances of the two effects, we find
\begin{equation}
\left(\frac{\delta\nu}{\nu}\right)^2_{ Total}\approx
\left(\frac{\delta\nu}{\nu}\right)^2_{ 
t}+\left(\frac{\delta\nu}{\nu}\right)^2_{g}
=\frac{\Omega_g^2}{\omega_m \Omega_t}+\frac{\Omega_t^2}{\omega_m^2}
\label{variance}
\end{equation}
The form of this function is plotted in Fig. 2.
(In the plot the numerical factors are reinserted.)
The total uncertainty in the clock rate
has a minimum at $\Omega_t^*=(\omega_m\Omega_g^2)^{1/3}$.
The corresponding value for the uncertainty in the clock rate is
\begin{equation}
\frac{\delta \nu}{\nu}\ge\frac{\Omega^*_t}{\omega_m}
\approx(\frac{\Omega_g}{\omega_m})^{\frac{2}{3}}
=\left(\frac{\lambdabar_c}{L_g}\right)^{\frac{2}{3}}
=\left(\frac{\hbar g}{m c^3}\right)^{\frac{2}{3}}.
\label{final}
\end{equation}
where $\lambdabar_c=\frac{\lambda_c}{2\pi}$ and
$\lambda_c=\frac{h}{mc}$ is the Compton wavelength of the clock.
We have thus derived a limit for the stability of clocks
in a gravitational field. 
As the gravitational field tends to $0$, so does $\Omega_t^*$ and we leave the Lamb-Dicke regime
and recover the Salecker-Wigner bound (Eq. ($6$) of Ref \cite{wignerq}). 
This is a fundamental limit on the stability of a clock with mass $m$. The 
limit is
considerably better than currently achievable stabilities. The importance
of the limit stems from the fact that it is a fundamental limit, telling
us something about the relation between, timekeeping, gravity and quantum 
mechanics.


Here we briefly describe possible systems
in which to test the quantum limits to timekeeping. The systems
in which the effects we describe are largest are the lighter clocks.
For example the Hydrogen maser could be compared with an Ytterbium
clock.
We can make an estimate of the uncertainty in clock rates
$\frac{\Delta T}{T}$ by inserting
the values of the relevant parameters for clocks of different types
in gravitational fields ranging from terrestrial to that of a Neutron 
star.
The results are displayed in table 1.
It is interesting to see that the present stabilities achieved so far are
in the range one part in $10^{18}$. Our analysis shows that there
is an ultimate achievable limit set by gravity.
The limit clearly depends on the gravitational field. The higher
the gravitational field, the worse the attainable stability.


We have shown that there is a fundamental quantum
limit to the stability achievable by a clock of mass $m$
in a gravitational field.
This limit comes from restrictions from the uncertainty principle on one
hand and the gravitational redshift on the other hand. Recent advances in
technology, especially in the domain of atomic physics have led to
highly stable clocks which are approaching the ultimate achievable limit.
We expect that these technological advances could be applied to
experimentally detect the limits we have derived.

Setting aside the experimental and technological aspects of this limit,
we note that it is a fundamental limit and cannot be violated by {\it any} 
clock. It gives
a limit to the stability of timekeeping because of
gravitational effects.  In some cases the well known quantum limit set by 
the width (or inverse lifetime) of
the excited state may be more stringent than our limit (\ref{final}) in a 
terrestrial gravitational field.
But in a sufficiently strong gravitational field the reverse will be true.
At a conceptual level,
Time is fundamental to General relativity as
Chance is to quantum mechanics. There is considerable current effort in
combining the two theories into a larger framework.
Much of the effort focuses on high energy physics and
attempts to unravel the microscopic structure of 
space-time\cite{smukhi,carlo,rafael}. In contrast, our effort here
is to understand low energy gravitational quantum physics. 
There are recent papers\cite{zych1,zych2,ss} which also address low energy
gravitational quantum physics. They go on to suggest relations between gravity
and decoherence\cite{decosup}. Such questions are of interest for 
further research but beyond the scope of the present paper.   
Given the enormous current interest in unifying gravity and
quantum mechanics, such low energy effects which may be testable
in a laboratory are of great interest.
\begin{figure}
\includegraphics[scale=0.5]{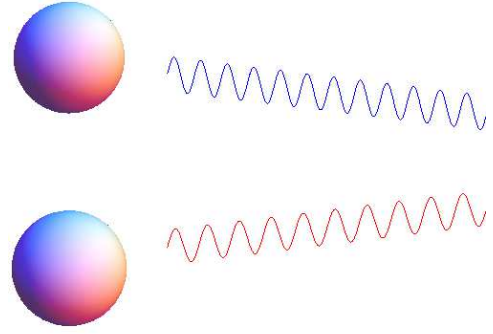}
\caption{\label{fig:epsart} Atomic Clock in a Gravitational 
Field 
$g$. Figure shows an atom in a trap emitting from two positions 
differing
in height. The signal received from the lower position appears red 
shifted
relative to the higher position. The two positions shown are within
the quantum uncertainty in vertical positions of the clock.}
\end{figure}

\begin{figure}
\includegraphics[scale=0.8]{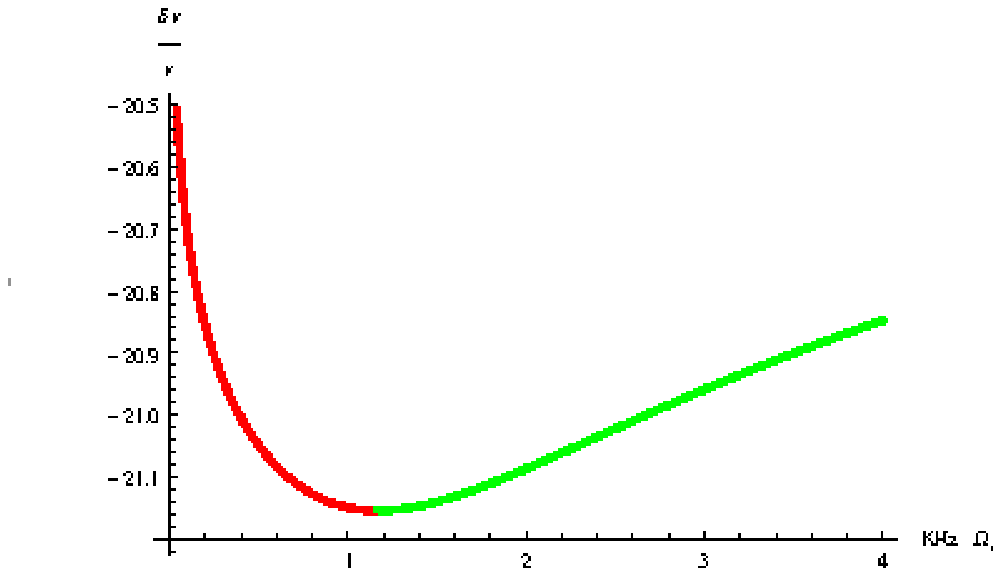}
\caption{\label{fig:epsart} Figure shows the logarithm (to base 
$10$) of
Fractional frequency instability $\frac{\delta \nu}{\nu}$ of a Hydrogen 
maser in a terrestrial gravitational field as a
function of the trap frequency $\Omega_t$ expressed in KiloHertz.
The major source of the uncertainty in the clock rate in
the red part of the curve is the gravitational redshift
and in the green part it is the time dilation effect. The optimal choice
for the trap frequency occurs at the join of the two parts of the curve,
where the fractional instability is $10^{-21}$.}
\end{figure}
\normalsize
\setlength{\textheight}{20cm}
\pagestyle{empty}
\begin{table}
  \begin{center}
    {\bf Table 1:~~~~~~~~Clock Stability}
\end{center}
\vspace*{0.5cm}
    \begin{tabular}{*{5}{|l}|}
      \hline
                          &           &                   &                           
& \\
      {\bf Clock Type } & $\nu_{c}$ & {\bf Atomic } & {\bf 
Gravitational } & {\bf Limit on }\\
                          &           &                   &                           
& \\
      {} & & {\bf  Mass} & {\bf Field} & {\bf Stability}\\
                          &           &                   &                           
& \\
      \hline
      Aluminium & $10^{15} H_{z}$ & 27 & Terrestrial & $10^{-22}$ 
\\*[0.1cm]
     \hline
      Aluminium & $10^{15} H_{z}$ & 27 & White Dwarf & $10^{-18}$ 
\\*[0.1cm]
     \hline
      Aluminium & $10^{15} H_{z}$ & 27 & Neutron Star& $10^{-14}$ 
\\*[0.1cm]
     \hline
      Hydrogen Maser & $10^{9} H_{z}$ & 1 & Terrestrial & $10^{-21}$ 
\\*[0.1cm]
     \hline
      Hydrogen Maser & $10^{9} H_{z}$ & 1 & White Dwarf & $10^{-17}$ 
\\*[0.1cm]
     \hline
      Hydrogen Maser & $10^{9} H_{z}$ & 1 & Neutron Star & $10^{-13}$ 
\\*[0.1cm]
     \hline
      Electron Spin  & $10^{13} H_{z}$ & 1/1836 & Terrestrial & 
$10^{-19}$ \\*[0.1cm]
      Precession &  &  &  & 
 \\*[0.1cm]
     \hline
      Electron Spin  & $10^{13} H_{z}$ & 1/1836 & White Dwarf & 
$10^{-15}$ \\*[0.1cm]
      Precession &  & & & 
 \\*[0.1cm]
     \hline
      Electron Spin  & $10^{13} H_{z}$ & 1/1836 & Neutron Star & 
$10^{-11}$ \\*[0.1cm]
      Precession &  & &  & 
\\*[0.1cm]
     \hline
      Neutron Spin & $10^{10} H_{z}$ & 1 & Terrestrial & 
$10^{-21}$ \\*[0.1cm]
    Precession  &  &  &  & 
 \\*[0.1cm]
     \hline
     Neutron Spin  & $10^{10} H_{z}$ & 1 & White Dwarf & 
$10^{-17}$ \\*[0.1cm]
     Precession  &  &  &  &  
\\*[0.1cm]
     \hline
      Neutron Spin  & $10^{10} H_{z}$ & 1 & Neutron Star & 
$10^{-13}$ \\*[0.1cm]
      Precession  &  &  & & 
$10^{-13}$ \\*[0.1cm]
      \hline
    \end{tabular}

\end{table}

\section{Acknowledgement}
It is a pleasure to thank John Barrow, Miguel Campiglia, Br\"ukner Caslav,
Fabio Costa, Nico Giulini, 
Bei-Lok Hu, H. S. Mani, Igor Pikovski, Hema Ramachandran,
Sadiq Rangwala, R. Sorkin, A. M. Srivastava and Magdalena Zych for discussions.

\bibliography{supurnareferences.bib}

\begin{thebibliography}{23}
\expandafter\ifx\csname natexlab\endcsname\relax\def\natexlab#1{#1}\fi
\expandafter\ifx\csname bibnamefont\endcsname\relax
  \def\bibnamefont#1{#1}\fi
\expandafter\ifx\csname bibfnamefont\endcsname\relax
  \def\bibfnamefont#1{#1}\fi
\expandafter\ifx\csname url\endcsname\relax
  \def\url#1{\texttt{#1}}\fi
\expandafter\ifx\csname urlprefix\endcsname\relax\def\urlprefix{URL }\fi
\providecommand*{\bibinfo}[2]{#2}
\providecommand*{\eprint}[1]{#1}
\providecommand*{\url}[1]{#1}
\begingroup\makeatletter
 \@temptokena{%
  \expandafter\ifx\csname citenamefont\endcsname\relax
   \DeclareRobustCommand\citenamefont{\@firstofone}%
   \global\let\citenamefont\citenamefont
   \global\expandafter\let\csname citenamefont \expandafter\endcsname\csname
  citenamefont \endcsname
  \fi
 }\if@filesw\immediate\write\@auxout{\the\@temptokena}\fi
\expandafter\endgroup\the\@temptokena

\bibitem[{\citenamefont{Sobel}(2007)}]{davasobel}
\bibinfo{author}{\bibfnamefont{D.}~\bibnamefont{Sobel}},
  \emph{\bibinfo{title}{Longitude: The True Story of a Lone Genius Who Solved
  the Greatest Scientific Problem of His Time}} (\bibinfo{publisher}{Walker and
  Company; Reprint edition (October 30, 2007)}, \bibinfo{year}{2007}).

\bibitem[{\citenamefont{Ramsey}(1983)}]{hist}
\bibinfo{author}{\bibfnamefont{N.~F.} \bibnamefont{Ramsey}},
  \bibinfo{journal}{Journal of Research of the National Bureau of Standards}
  \textbf{\bibinfo{volume}{88}}(\bibinfo{number}{5}), \bibinfo{pages}{301}
  (\bibinfo{year}{1983}).

\bibitem[{\citenamefont{Winkler and Van~Flandern}(1977)}]{astro}
\bibinfo{author}{\bibfnamefont{G.~M.~R.} \bibnamefont{Winkler}}
  \bibnamefont{and} \bibinfo{author}{\bibfnamefont{T.~C.}
  \bibnamefont{Van~Flandern}}, \bibinfo{journal}{The Astronomical Journal}
  \textbf{\bibinfo{volume}{82}}, \bibinfo{pages}{84} (\bibinfo{year}{1977}).

\bibitem[{\citenamefont{Lorimer and Kramer}(2005)}]{handbook}
\bibinfo{author}{\bibfnamefont{D.~R.} \bibnamefont{Lorimer}} \bibnamefont{and}
  \bibinfo{author}{\bibfnamefont{M.}~\bibnamefont{Kramer}},
  \emph{\bibinfo{title}{Handbook of Pulsar Astronomy}}
  (\bibinfo{publisher}{Cambridge University Press}, \bibinfo{year}{2005}).

\bibitem[{\citenamefont{Thomson and Tait}(1879)}]{kelvin}
\bibinfo{author}{\bibfnamefont{W.}~\bibnamefont{Thomson}} \bibnamefont{and}
  \bibinfo{author}{\bibfnamefont{P.~G.} \bibnamefont{Tait}},
  \emph{\bibinfo{title}{Treatise on Natural Philosophy}}
  (\bibinfo{publisher}{Cambridge University Press}, \bibinfo{year}{1879}).

\bibitem[{\citenamefont{Kusch} \emph{et~al.}(1940)\citenamefont{Kusch, Millman,
  and Rabi}}]{rabi}
\bibinfo{author}{\bibfnamefont{P.}~\bibnamefont{Kusch}},
  \bibinfo{author}{\bibfnamefont{S.}~\bibnamefont{Millman}}, \bibnamefont{and}
  \bibinfo{author}{\bibfnamefont{I.~I.} \bibnamefont{Rabi}},
  \bibinfo{journal}{Phys. Rev.} \textbf{\bibinfo{volume}{57}},
  \bibinfo{pages}{765} (\bibinfo{year}{1940}),
  \urlprefix\url{http://link.aps.org/doi/10.1103/PhysRev.57.765}.

\bibitem[{\citenamefont{Chou} \emph{et~al.}(2010)\citenamefont{Chou, Hume,
  Rosenbland, and Wineland}}]{wineland}
\bibinfo{author}{\bibfnamefont{C.}~\bibnamefont{Chou}},
  \bibinfo{author}{\bibfnamefont{D.}~\bibnamefont{Hume}},
  \bibinfo{author}{\bibfnamefont{T.}~\bibnamefont{Rosenbland}},
  \bibnamefont{and} \bibinfo{author}{\bibfnamefont{D.}~\bibnamefont{Wineland}},
  \bibinfo{journal}{Science} \textbf{\bibinfo{volume}{329}},
  \bibinfo{pages}{1630} (\bibinfo{year}{2010}).

\bibitem[{\citenamefont{Hinkley} \emph{et~al.}(2013)\citenamefont{Hinkley,
  Sherman, Phillips, Schioppo, Lemke, Beloy, Pizzocaro, Oates, and
  Ludlow}}]{nist}
\bibinfo{author}{\bibfnamefont{N.}~\bibnamefont{Hinkley}},
  \bibinfo{author}{\bibfnamefont{J.~A.} \bibnamefont{Sherman}},
  \bibinfo{author}{\bibfnamefont{N.~B.} \bibnamefont{Phillips}},
  \bibinfo{author}{\bibfnamefont{M.}~\bibnamefont{Schioppo}},
  \bibinfo{author}{\bibfnamefont{N.~D.} \bibnamefont{Lemke}},
  \bibinfo{author}{\bibfnamefont{K.}~\bibnamefont{Beloy}},
  \bibinfo{author}{\bibfnamefont{M.}~\bibnamefont{Pizzocaro}},
  \bibinfo{author}{\bibfnamefont{C.~W.} \bibnamefont{Oates}}, \bibnamefont{and}
  \bibinfo{author}{\bibfnamefont{A.~D.} \bibnamefont{Ludlow}},
  \bibinfo{journal}{Science} \textbf{\bibinfo{volume}{341}},
  \bibinfo{pages}{1215} (\bibinfo{year}{2013}).

\bibitem[{\citenamefont{Salecker and Wigner}(1958)}]{wignerq}
\bibinfo{author}{\bibfnamefont{H.}~\bibnamefont{Salecker}} \bibnamefont{and}
  \bibinfo{author}{\bibfnamefont{E.~P.} \bibnamefont{Wigner}},
  \bibinfo{journal}{Phys. Rev.} \textbf{\bibinfo{volume}{109}},
  \bibinfo{pages}{571} (\bibinfo{year}{1958}),
  \urlprefix\url{http://link.aps.org/doi/10.1103/PhysRev.109.571}.

\bibitem[{\citenamefont{Wigner}(1957)}]{wignerrmp}
\bibinfo{author}{\bibfnamefont{E.~P.} \bibnamefont{Wigner}},
  \bibinfo{journal}{Rev. Mod. Phys.} \textbf{\bibinfo{volume}{29}},
  \bibinfo{pages}{255} (\bibinfo{year}{1957}),
  \urlprefix\url{http://link.aps.org/doi/10.1103/RevModPhys.29.255}.

\bibitem[{\citenamefont{Barrow}(1996)}]{barrow}
\bibinfo{author}{\bibfnamefont{J.~D.} \bibnamefont{Barrow}},
  \bibinfo{journal}{Phys. Rev. D} \textbf{\bibinfo{volume}{54}},
  \bibinfo{pages}{6563} (\bibinfo{year}{1996}),
  \urlprefix\url{http://link.aps.org/doi/10.1103/PhysRevD.54.6563}.

\bibitem[{\citenamefont{Dicke}(1953)}]{dicke}
\bibinfo{author}{\bibfnamefont{R.~H.} \bibnamefont{Dicke}},
  \bibinfo{journal}{Phys. Rev.} \textbf{\bibinfo{volume}{89}},
  \bibinfo{pages}{472} (\bibinfo{year}{1953}),
  \urlprefix\url{http://link.aps.org/doi/10.1103/PhysRev.89.472}.

\bibitem[{\citenamefont{Cohen-Tannoudji and
  {Gu{\'e}ry}-Odelin}(2011)}]{cohentannoud}
\bibinfo{author}{\bibfnamefont{C.}~\bibnamefont{Cohen-Tannoudji}}
  \bibnamefont{and}
  \bibinfo{author}{\bibfnamefont{D.}~\bibnamefont{{Gu{\'e}ry}-Odelin}},
  \emph{\bibinfo{title}{Advances in Atomic Physics:An Overview}}
  (\bibinfo{publisher}{World Scientific}, \bibinfo{year}{2011}).

\bibitem[{\citenamefont{Schilpp}(1970)}]{solvay}
\bibinfo{author}{\bibfnamefont{P.}~\bibnamefont{Schilpp}},
  \emph{\bibinfo{title}{Albert Einstein: Philosopher-scientist}}
  (\bibinfo{publisher}{Open Court Publishing Co, US}, \bibinfo{year}{1970}).

\bibitem[{\citenamefont{Campbell} \emph{et~al.}(2012)\citenamefont{Campbell,
  Radnaev, Kuzmich, Dzuba, Flambaum, and Derevianko}}]{nuclearclocks}
\bibinfo{author}{\bibfnamefont{C.~J.} \bibnamefont{Campbell}},
  \bibinfo{author}{\bibfnamefont{A.~G.} \bibnamefont{Radnaev}},
  \bibinfo{author}{\bibfnamefont{A.}~\bibnamefont{Kuzmich}},
  \bibinfo{author}{\bibfnamefont{V.~A.} \bibnamefont{Dzuba}},
  \bibinfo{author}{\bibfnamefont{V.~V.} \bibnamefont{Flambaum}},
  \bibnamefont{and}
  \bibinfo{author}{\bibfnamefont{A.}~\bibnamefont{Derevianko}},
  \bibinfo{journal}{Phys. Rev. Lett.} \textbf{\bibinfo{volume}{108}},
  \bibinfo{pages}{120802} (\bibinfo{year}{2012}),
  \urlprefix\url{http://link.aps.org/doi/10.1103/PhysRevLett.108.120802}.

\bibitem[{\citenamefont{{M{\"o}ssbauer}}(1958)}]{mossbauer}
\bibinfo{author}{\bibfnamefont{R.~L.} \bibnamefont{{M{\"o}ssbauer}}},
  \bibinfo{journal}{Zeitschrift fur Physik} \textbf{\bibinfo{volume}{151}},
  \bibinfo{pages}{124} (\bibinfo{year}{1958}).

\bibitem[{\citenamefont{Mukhi}(2011)}]{smukhi}
\bibinfo{author}{\bibfnamefont{S.}~\bibnamefont{Mukhi}},
  \bibinfo{journal}{Classical and Quantum Gravity}
  \textbf{\bibinfo{volume}{28}}(\bibinfo{number}{15}), \bibinfo{pages}{153001}
  (\bibinfo{year}{2011}).

\bibitem[{\citenamefont{Rovelli}(2011)}]{carlo}
\bibinfo{author}{\bibfnamefont{C.}~\bibnamefont{Rovelli}},
  \bibinfo{journal}{Classical and Quantum Gravity}
  \textbf{\bibinfo{volume}{28}}(\bibinfo{number}{15}), \bibinfo{pages}{153002}
  (\bibinfo{year}{2011}).

\bibitem[{\citenamefont{Sorkin}(2009)}]{rafael}
\bibinfo{author}{\bibfnamefont{R.~D.} \bibnamefont{Sorkin}},
  \bibinfo{journal}{J.Phys.Conf.Ser} \textbf{\bibinfo{volume}{174}},
  \bibinfo{pages}{012018} (\bibinfo{year}{2009}),
  \urlprefix\url{arXiv:0910.0673}.

\bibitem[{\citenamefont{Zych} \emph{et~al.}(2011)\citenamefont{Zych, Costa,
  Pikovski, and Brukner}}]{zych1}
\bibinfo{author}{\bibfnamefont{M.}~\bibnamefont{Zych}},
  \bibinfo{author}{\bibfnamefont{F.}~\bibnamefont{Costa}},
  \bibinfo{author}{\bibfnamefont{I.}~\bibnamefont{Pikovski}}, \bibnamefont{and}
  \bibinfo{author}{\bibfnamefont{C.}~\bibnamefont{Brukner}},
  \bibinfo{journal}{Nat Commun} \textbf{\bibinfo{volume}{2}}
  (\bibinfo{year}{2011}), \urlprefix\url{http://dx.doi.org/10.1038/ncomms1498}.

\bibitem[{\citenamefont{Pikovski} \emph{et~al.}(2013)\citenamefont{Pikovski,
  Zych, Costa, , and Brukner}}]{zych2}
\bibinfo{author}{\bibfnamefont{I.}~\bibnamefont{Pikovski}},
  \bibinfo{author}{\bibfnamefont{M.}~\bibnamefont{Zych}},
  \bibinfo{author}{\bibfnamefont{F.}~\bibnamefont{Costa}}, \bibinfo{author}{~},
  \bibnamefont{and} \bibinfo{author}{\bibfnamefont{C.}~\bibnamefont{Brukner}}
  (\bibinfo{year}{2013}), \urlprefix\url{http://arxiv.org/abs/1311.1095}.

\bibitem[{\citenamefont{Sinha and Samuel}(2011)}]{ss}
\bibinfo{author}{\bibfnamefont{S.}~\bibnamefont{Sinha}} \bibnamefont{and}
  \bibinfo{author}{\bibfnamefont{J.}~\bibnamefont{Samuel}},
  \bibinfo{journal}{Classical and Quantum Gravity}
  \textbf{\bibinfo{volume}{28}}(\bibinfo{number}{14}), \bibinfo{pages}{145018}
  (\bibinfo{year}{2011}).

\bibitem[{\citenamefont{Sinha}(1997)}]{decosup}
\bibinfo{author}{\bibfnamefont{S.}~\bibnamefont{Sinha}},
  \bibinfo{journal}{Physics Letters A} \textbf{\bibinfo{volume}{228}},
  \bibinfo{pages}{1} (\bibinfo{year}{1997}).

\end{thebibliography}

\end{document}